\documentclass[aps,pra,showpacs,amsmath,amssymb,amsfonts,superscriptaddress]{revtex4-2}
\usepackage{hyperref}
\usepackage{dcolumn}
\usepackage{times}
\usepackage{graphicx}
\usepackage{verbatim}
\usepackage{bm}

\usepackage{ragged2e}  
\usepackage{xcolor}
\usepackage{hyperref}
\usepackage{hhline}
\usepackage{caption}
\usepackage{subcaption}
\usepackage{comment}
\usepackage{siunitx}
\usepackage{float}
\usepackage{enumerate}
\usepackage{bbm}
\usepackage{tikz,pgfplots}
\usepgfplotslibrary{fillbetween}
\usetikzlibrary{backgrounds}
\usetikzlibrary{patterns}

\hypersetup{
    colorlinks=true,
    linkcolor=blue,
    citecolor=blue,
    urlcolor=blue
}

\begin{document}

\title{Collective Dissipation and Parameter Sensitivity in Trapped Ions Coupled to a Common Thermal Reservoir}

\author{C. F. P. Avalos}
\email{cfavalos@ifi.unicamp.br}
\affiliation{Instituto de F\'\i sica  Gleb Wataghin, Universidade Estadual de Campinas, 13083-859, Campinas, SP, Brazil} 

\author{G. A. Prataviera }
\email{prataviera@usp.br}
\affiliation{Departamento de Administra\c{c}\~ao, FEA-RP, Universidade de S\~{a}o Paulo, 14040-905, Ribeir\~{a}o Preto, SP, Brazil}

\author{M. C. de Oliveira}
\email{marcos@ifi.unicamp.br}
\affiliation{Instituto de F\'\i sica  Gleb Wataghin, Universidade Estadual de Campinas, 13083-859, Campinas, SP, Brazil} 

\begin{abstract}
We investigate the dynamics of two trapped ions interacting with a common thermal reservoir, focusing on how cross-correlated dissipation influences heating, steady-state behavior, and parameter sensitivity. Starting from a microscopic system--reservoir model, we derive the corresponding Heisenberg--Langevin equations and show that reservoir-induced correlations generate collective decay channels and, when the cross-damping rate matches the local damping, a decoherence-free normal mode that preserves memory of the initial excitations. Using the Fisher information associated with motional population measurements,  we identify the parameter regimes in which cross-damping enhances the estimability of both system and reservoir properties. For nonclassical initial states, we also show that reservoir-mediated correlations can generate or maintain entanglement, with the strongest effects occurring near the decoherence-free condition. 
\end{abstract}

\date{\today}

\maketitle

\section{Introduction}

Trapped ions are among the most versatile and precisely controllable platforms
for quantum information processing, quantum simulation, and precision
metrology~\cite{Brown2011,Harlander2011,SaNeto2022}. Their internal electronic
states can be addressed with high fidelity, and their quantized motional modes
offer long coherence times and flexible engineering of interactions. These
features have enabled state-of-the-art demonstrations in quantum logic,
entanglement generation, and ultra-sensitive force and field sensing.

Despite this high degree of control, trapped-ion performance remains limited by
motional heating and decoherence. Fluctuating electromagnetic fields near trap
surfaces, technical noise, and coupling to uncontrolled environmental degrees of
freedom can inject energy into the motional modes, degrade nonclassical
resources, and ultimately constrain gate fidelity and metrological
sensitivity~\cite{SaNeto2022,Mandel1995}. 
In the standard description, each ion
interacts with an effectively independent reservoir, so dissipation acts
locally with motional states relaxing individually through independent decay channels towards thermal equilibrium,
suppressing coherence, squeezing, and entanglement.

In many experimental architectures, however, ions do not couple to independent
noise sources. Shared trap electrodes, common voltage supplies, and the finite
correlation length of fluctuating surface potentials lead to situations in which
environmental fluctuations are at least partially correlated across the ion
array. In this collective-dissipation regime, cross-damping terms modify
the relaxation dynamics and give rise to qualitatively new phenomena. These
include superradiant and subradiant decay channels, modified steady-state
occupations, and the possibility of a decoherence-free subspace (DFS) when the
cross-damping rate matches the local damping~\cite{Braun2002,Benatti2003,Paz2008}.
Moreover, a common reservoir can mediate correlations or even generate
entanglement between otherwise uncoupled degrees of freedom, as shown for both
two-level systems~\cite{Braun2002,Benatti2003,Romanelli2022,An2007} and bosonic
modes~\cite{Wolf2011,Paz2008,Prauzner2004,DeCastro2008}.

Although these effects are well documented individually, a unified analysis that
connects a {microscopically derived} cross-damped Langevin description with
{quantitative measures of parameter sensitivity and entanglement} remains
largely unexplored. The present work aims to provide such a comprehensive
treatment for two Coulomb-coupled trapped ions interacting with a shared thermal
reservoir. Our approach combines a fully analytic derivation of the dissipative
normal modes with an information-theoretic analysis of estimability and a
Gaussian-state entanglement study.
Related analyses of non-classical motional dynamics in trapped-ion systems have been developed in the unitary, non-adiabatic regime, where squeezing and P-nonclassicality emerge as resources for quantum thermodynamics and sensing \cite{AvalosNJP2025}.

Beyond consolidating established aspects of collective decay and decoherence-free subspaces, our work advances the field in three key directions. First, we provide a microscopic and fully analytic treatment of cross-damped Langevin dynamics for Coulomb-coupled ions, enabling an exact identification of collective decay channels and their dependence on reservoir correlations. Second, we connect this dynamical structure to information-theoretic performance by computing the full classical Fisher information matrix for experimentally accessible phonon-number measurements. This reveals how correlated dissipation can enhance or even preserve the estimability of system and reservoir parameters—an aspect not addressed in previous analyses of trapped-ion thermometry or collective damping. Third, we demonstrate that the same mechanism that enhances metrological performance also governs the creation and persistence of Gaussian entanglement, providing a unified physical picture of how reservoir correlations act simultaneously as a generator and stabilizer of nonclassical motional states. These results establish collective dissipation not only as a decoherence mechanism but as a tunable resource for precision sensing and quantum-state engineering in trapped-ion platforms.

The paper is organized as follows. Section~II presents the two-ion Hamiltonian
and summarizes the approximations used. Section~III introduces the interaction
with a common thermal reservoir and derives the corresponding
Heisenberg--Langevin equations. Section~IV analyzes the resulting dynamics,
including collective decay rates, heating behavior, and the emergence of a
decoherence-free subspace. In Sec.~V we compute the Fisher information for
population measurements and discuss the distinct estimation regimes that emerge
for different correlation strengths. Section~VI examines the entanglement
dynamics of nonclassical initial states under collective dissipation. Our
conclusions are presented in Sec.~VII.

\section{Two-ion system}

We consider two ions of mass $m$ and charge $q$ confined in a linear Paul trap and separated by a distance $d$. The ions oscillate around their equilibrium positions and interact through the Coulomb repulsion. We focus on the longitudinal motional degree of freedom along the ion--ion axis, for which the system Hamiltonian can be written as~\cite{Brown2011,Harlander2011,SaNeto2022}
\begin{equation}
H = \sum_{j=1}^2 \left( \frac{p_j^2}{2m} + \frac{1}{2} m \omega^2 x_j^2 \right) + H_{\mathrm{C}},
\label{H}
\end{equation}
where $\omega$ is the trapping frequency, and  
\begin{equation}
H_{\mathrm{C}} = \frac{k q^2}{d + x_2 - x_1},
\label{Hc}
\end{equation}
with $k = (4\pi\varepsilon_0)^{-1}$.

For small displacements, $|x_2 - x_1| \ll d$, we expand Eq.~(\ref{Hc}) to second order as
\begin{equation}
H_{\mathrm{C}} \approx \frac{k q^2}{d} \left[ 1 - \frac{x_2 - x_1}{d} + \frac{(x_2 - x_1)^2}{d^2} \right].
\label{Hcp}
\end{equation}

In Eq. \ref{Hcp}, the constant term contributes only an overall energy offset and can be ignored, while the linear term merely shifts the equilibrium positions and the quadratic terms modify the trap frequency and generate an effective bilinear coupling between the ions.

Introducing bosonic operators  
\begin{equation}
x_j = \sqrt{\frac{\hbar}{2m\omega}} (a_j + a_j^\dagger), 
\qquad
p_j = i\sqrt{\frac{\hbar m\omega}{2}} (a_j^\dagger - a_j),
\end{equation}
the Hamiltonian becomes
\begin{align}
H &= \hbar \omega_0 (a_1^\dagger a_1 + a_2^\dagger a_2)
+  \sqrt{\frac{\hbar k^2 q^4}{2m\omega d^4}}(a_{1}+a_{1}^{\dagger}-a_{2}-a_{2}^{\dagger})\nonumber\\
&\quad +\frac{\hbar k q^2}{2m\omega d^3} \left(a_1^2 + a_1^{\dagger 2} + a_2^2 + a_2^{\dagger 2}\right)
\nonumber\\
&\quad - \frac{\hbar k q^2}{m\omega d^3} 
\left(a_1 a_2^\dagger + a_2 a_1^\dagger + a_1 a_2 + a_1^\dagger a_2^\dagger\right)
,
\end{align}
where $\omega_0 = \omega + k q^2/(m\omega d^3)$  are the renormalized ions frequencies, which includes the small frequency shift due to the Coulomb interaction.

Moving to the interaction picture with respect to  
\[
H_0 = \hbar\omega_{0}(a_1^\dagger a_1 + a_2^\dagger a_2),
\]
and applying the rotating-wave approximation to neglect terms oscillating at  $\pm \omega_0$, and $\pm 2\omega_0$ (which average out on the timescales of interest), we are left with the dominant bilinear coupling
\begin{equation}
\tilde{H} = \hbar \Omega \left( a_1 a_2^\dagger + a_1^\dagger a_2 \right),
\label{H12}
\end{equation}
where $\Omega = -k q^2/(m \omega d^3)$ is the effective coupling strength.

Equation~(\ref{H12}) describes a beam-splitter--type interaction between the motional modes. Such a Hamiltonian does not generate entanglement from classical initial states, but can generate or exchange entanglement when at least one of the initial states is nonclassical. In the absence of environmental coupling, the dynamics are fully coherent and consist of periodic energy exchange between the two ions. As we show in the next section, coupling to a common environment qualitatively modifies this behavior through the appearance of cross-damping terms, which play a central role in the emergence of collective decay and decoherence-free subspaces.

\section{Damping and heating}

We now incorporate dissipation by coupling the ions to a common thermal reservoir. The environment is modeled as a collection of bosonic modes with frequencies $\omega_k$, which couple linearly to the motional annihilation operators of both ions. The full Hamiltonian reads
\begin{align}
H &= \sum_{j=1}^2 \hbar\omega_0 a_j^\dagger a_j 
+  \hbar\Omega \left(a_1 a_2^\dagger + a_1^\dagger a_2\right)\nonumber\\
&\quad + \sum_k \hbar\omega_k b_k^\dagger b_k  + \sum_{j=1}^2 \sum_k \hbar\left(g_{jk} b_k a_j^\dagger + g_{jk}^* b_k^\dagger a_j\right),
\label{ham_total}
\end{align}
 where $g_{jk}$ denotes the coupling strength between ion $j$ and bath mode $k$, and the zero-point energies have been omitted, as they only add a constant shift to the Hamiltonian and do not affect the dynamics. The operators $a_j$ and $b_k$ annihilate excitations of the ions and the reservoir, respectively

The Heisenberg equations of motion are
\begin{align}
\dot{a}_j &= -i\omega_0 a_j - i\Omega a_l - i\sum_k g_{jk} b_k, 
\label{eq_a_j} \\
\dot{b}_k &= -i\omega_k b_k - i\sum_{j=1}^2 g_{jk}^* a_j,
\label{eq_b_k}
\end{align}
with $j\neq l =1,2$. Solving Eq.~(\ref{eq_b_k}) formally and substituting into Eq.~(\ref{eq_a_j}) yields
\begin{align}
\dot{a}_j &= -i\omega_0 a_j - i\Omega a_l - i\sum_k g_{jk} b_k(0)e^{-i\omega_k t}
\nonumber\\
&\quad - \sum_k |g_{jk}|^2 \int_0^t \! d\tau\, a_j(t-\tau)e^{-i\omega_k\tau}\nonumber\\
&\quad - \sum_k g_{jk} g_{lk}^* \int_0^t \! d\tau\, a_l(t-\tau)e^{-i\omega_k\tau}.
\label{eq_before_markov}
\end{align}

The first two terms describe the system's coherent dynamics. The third term defines a noise operator,
\begin{equation}
F_j(t) = -i\sum_k g_{jk} b_k(0)e^{-i\omega_k t},
\end{equation}
while the last two terms represent dissipation generated by the reservoir. To evaluate these integrals, we replace the discrete sum over bath modes by a continuous distribution with density $\eta_{jl}(\omega)$ and define the cross spectral density
\begin{equation}
W_{jl}(\omega) = \eta_{jl}(\omega) g_j(\omega) g_l^*(\omega),
\end{equation}
which satisfies~\cite{Mandel1995}
\begin{equation}
W_{jl} = W_{lj}^*, \qquad |W_{jl}| \leq \sqrt{W_{jj} W_{ll}}.
\end{equation}

The generalized Wiener--Khinchin theorem gives the reservoir autocorrelation function
\begin{equation}
\Gamma_{jl}(\tau)= \int_{-\omega_0}^{\infty} W_{jl}(\omega_0+\omega') e^{-i\omega'\tau} d\omega'.
\end{equation}
Under the Markov approximation, $W_{jl}(\omega)$ varies slowly around $\omega_0$ so that $W_{jl}(\omega_0+\omega')\approx W_{jl}(\omega_0)$, yielding
\begin{equation}
\Gamma_{jl}(\tau)= 2\pi W_{jl}(\omega_0)\delta(\tau).
\end{equation}
Inserting this into Eq.~(\ref{eq_before_markov}) gives the dissipative contribution
\begin{equation}
\sum_k g_{jk}g_{lk}^* \int_0^t a_l(t-\tau)e^{-i\omega_k\tau} \, d\tau
\;\rightarrow\; \frac{\gamma_{jl}}{2} a_l(t),
\end{equation}
where the damping rates are
\begin{equation}
\gamma_{jl} = 2\pi \eta_{jl}(\omega_0) g_j(\omega_0) g_l^*(\omega_0).
\end{equation}
The Cauchy--Schwarz inequality implies
\begin{equation}
\gamma_{jl} \le \sqrt{\gamma_{jj}\gamma_{ll}},
\label{ineq}
\end{equation}
showing that the cross-damping cannot exceed the local dissipation.

The noise operators satisfy, for a bath with mean thermal occupation $\bar{N}$,
\begin{align}
\langle F_j(t)\rangle &=0, \\
\langle F_j(t) F_l(t')\rangle &=0, \\
\langle F_j^\dagger(t) F_l(t')\rangle &= \bar{N}\,\gamma_{jl}\,\delta(t-t'), \\
\langle F_j(t) F_l^\dagger(t')\rangle &= (\bar{N}+1)\,\gamma_{jl}\,\delta(t-t').
\end{align}

Collecting all terms, we obtain the Heisenberg--Langevin equations
\begin{equation}
\dot{a}_j(t) = -\left(i\omega_0 + \frac{\gamma_{jj}}{2}\right)a_j(t)
 - \left(i\Omega + \frac{\gamma_{jl}}{2}\right)a_l(t)
 + F_j(t),
\label{HL_final}
\end{equation}
for $ j,l=1,2,$, showing that the common reservoir induces not only local damping $\gamma_{jj}$ but also cross-damping $\gamma_{jl}$. The latter term arises from interference between decay channels and is responsible for the collective dynamics discussed in the next section.

\section{Dynamics and equilibrium}

The Heisenberg--Langevin equations obtained in Sec.~III describe the coupled dynamics of the ions under both coherent tunneling and dissipative interactions with the common reservoir. To analyze these equations, we first consider the general case without imposing any symmetry on the damping matrix, and only afterwards apply the physically motivated assumption that the ions are identical and couple equivalently to the bath. 

Introducing the vector $\hat{\mathbf{a}}=(\hat{a}_{1},\hat{a}_{2})^{T}$ and the Langevin noise vector,  $\mathbf{F}(t)= \left( F_{1}(t)\ F_{2}(t)\right)^{T}$, the equations of motion can be written as
\begin{equation}
    \dot{\hat{\mathbf{a}}}(t) = -i\mathbf{M}\hat{\mathbf{a}}(t) + \mathbf{F}(t),
    \label{eq:generalEOM}
\end{equation}
where the drift matrix
\begin{equation}
    \mathbf{M} =
    \begin{pmatrix}
        \omega_{0} - i\gamma_{11}/2 & \Omega - i\gamma_{12}/2 \\
        \Omega - i\gamma_{21}/2 & \omega_{0}-i\gamma_{22}/2
    \end{pmatrix}
    \label{eq:M_general}
\end{equation}
contains both the coherent coupling $\Omega$ and all dissipative channels. The formal solution of Eq.~\eqref{eq:generalEOM} is
\begin{equation}
\hat{\mathbf{a}}(t)
= e^{-i\mathbf{M}t}\hat{\mathbf{a}}(0)
+ \int_{0}^{t} e^{-i\mathbf{M}(t-t')}\mathbf{F}(t')\,dt' .
\label{eq:formalSolution}
\end{equation}

The matrix exponential $e^{-i\mathbf{M}t}$ can be expressed analytically by diagonalizing $\mathbf{M}$. The eigenvalues are

\begin{equation}
\lambda_{\pm}
= \omega_0 - i\frac{\gamma_{11} + \gamma_{22}}{4}
\pm \sqrt{\Delta},
\label{eq:eigen_general}
\end{equation}
with
\begin{equation*}
\Delta=\left( \Omega - i \frac{\gamma_{12}}{2} \right)
           \left( \Omega - i \frac{\gamma_{21}}{2} \right)
           - \left( \frac{\gamma_{11} - \gamma_{22}}{4} \right)^2, 
\end{equation*}
showing explicitly how non-symmetric dissipation affects the normal-mode dynamics. This expression is the most general form and reduces to simpler structures only after physical symmetry is imposed. Since it was assumed identical ions subjected to the same environment, it is physically reasonable to assume $\gamma_{12}=\gamma_{21}$, and $\gamma_{11}=\gamma_{22}\equiv \gamma$, consistent with the positivity condition $\gamma_{12} \leq \gamma $ obtained in Sec.~III. Under these conditions, the eigenvalues simplify to
\begin{equation}
\lambda_{\pm}
= \omega_{0}\pm \Omega -i\frac{\gamma\pm \gamma_{12}}{2},
\label{eq:eigen_symmetric}
\end{equation}
which reveal the emergence of collective decay channels. Introducing the symmetric and antisymmetric modes,
\begin{equation}
A = \frac{a_{1}+a_{2}}{\sqrt{2}},
\qquad
B = \frac{a_{1}-a_{2}}{\sqrt{2}},
\label{eq:normalModes}
\end{equation}
the equations of motion become
\begin{align}
\dot{A}(t) &= -\left[i(\omega_{0}+\Omega) + \frac{\Gamma_{+}}{2}\right]A(t) + F_{+}(t),
\\[1mm]
\dot{B}(t) &= -\left[i(\omega_{0}-\Omega) + \frac{\Gamma_{-}}{2}\right]B(t) + F_{-}(t),
\end{align}
where the collective decay rates are
\begin{equation}
\Gamma_{\pm} = \gamma \pm \gamma_{12},
\label{eq:collective_rates}
\end{equation}
and $F_{\pm} (t)$ are effective langevin forces.

Equation~\eqref{eq:collective_rates} demonstrates that the common reservoir induces superradiant and subradiant decay channels. When $\gamma_{12}=0$ the modes decay at the same rate $\gamma$,  which is physically equivalent to the case of coupling to two independent local baths. As $\gamma_{12}$ increases, the symmetric mode $A$ decays faster, while the antisymmetric mode $B$ decays more slowly. At the boundary of complete reservoir correlation, $\gamma_{12} = \gamma$ we obtain
\begin{equation}
\Gamma_{+}=2\gamma,
\qquad
\Gamma_{-}=0,
\end{equation}
and the antisymmetric mode $B$ becomes immune to dissipation—a decoherence-free subspace (DFS). In this regime, all excitations stored in $B$ persist indefinitely, and the system retains a memory of its initial state even in the presence of the thermal reservoir.

Using the general solution \eqref{eq:formalSolution}, and assuming the ions initially not correlated, the mean excitation of ion $j$ takes the compact form
\begin{align}
n_{j}(t) &= e^{-\gamma t}\!\left[n_{j}(0)\cos^{2}(\Omega t) + n_{l}(0)\sin^{2}(\Omega t)\right]
\nonumber\\
&\quad + \frac{1}{2}e^{-\gamma t}\!\left[n_{1}(0)+n_{2}(0)\right]\big(\cosh(\gamma_{12}t)-1\big)
\nonumber\\
&\quad+ \bar{N}\left[1-e^{-\gamma t}\cosh(\gamma_{12}t)\right].
\label{eq:population_full}
\end{align}

\begin{figure}[t]
  \centering
  \begin{subfigure}[t]{0.48\textwidth}
    \centering
    \begin{tikzpicture}
      \node[inner sep=0pt] (imgA) {\includegraphics[width=\linewidth]{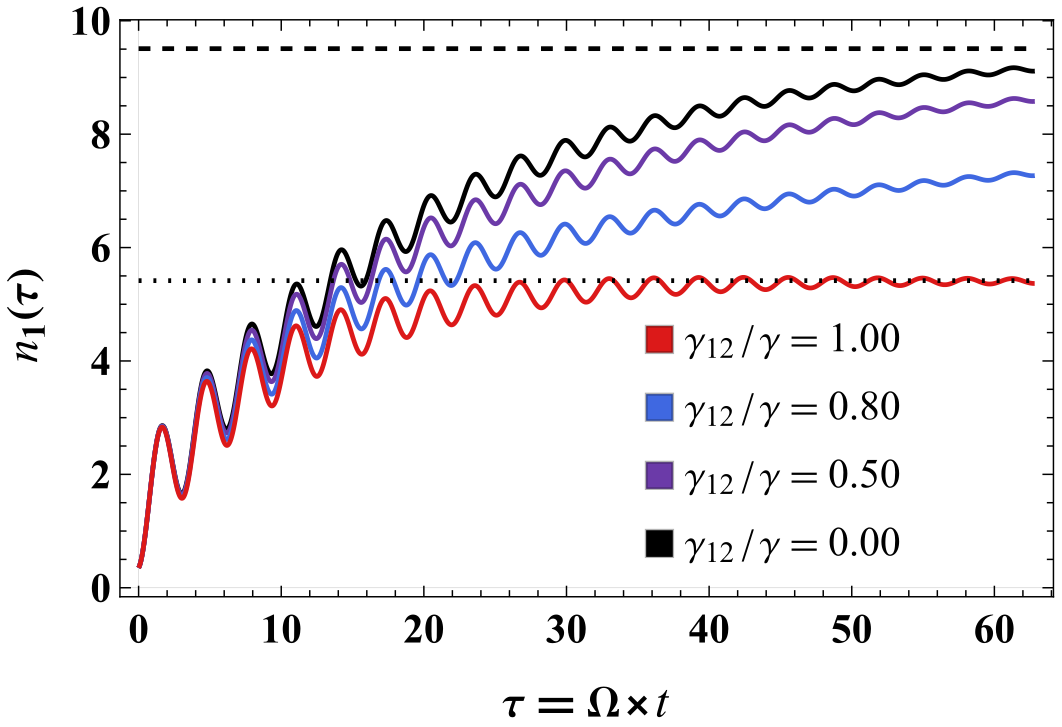}};
      \node[anchor=north west, font=\bfseries, xshift=4pt, yshift=-4pt] at (imgA.north west) {(a)};
    \end{tikzpicture}
  \end{subfigure}
  \hfill
  \begin{subfigure}[t]{0.48\textwidth}
    \centering
    \begin{tikzpicture}
      \node[inner sep=0pt] (imgB) {\includegraphics[width=\linewidth]{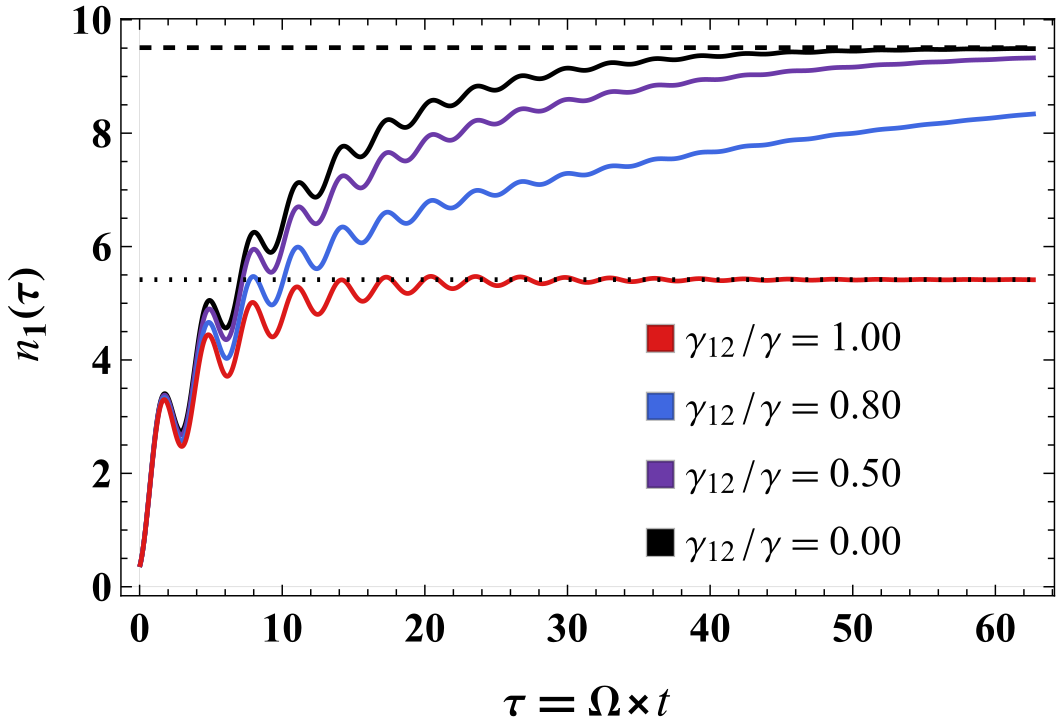}};
      \node[anchor=north west, font=\bfseries, xshift=4pt, yshift=-4pt] at (imgB.north west) {(b)};
    \end{tikzpicture}
  \end{subfigure}
 \caption{\justifying (Color online) Time evolution of the mean phonon number of ion~1 for several
 values of the normalized cross-damping rate $\gamma_{12}/\gamma$. The curves, from top to bottom, correspond to $\gamma_{12}/\gamma = 0.00, 0.50, 0.80, 1.00$. The ions are initially prepared in thermal state with $\bar{n}_1=0.35$ and $\bar{n}_2=2.3$. 
 The black dashed line marks the thermal occupation $\bar{N}$ of the reservoir for $\hbar \omega_{0}/k_{B}T = 0.10$,
 and the black dotted line marks the steady-state population in the
 fully correlated limit $\gamma_{12}=\gamma$.
 Panels (a) and (b) correspond to $\gamma/\Omega = 0.05$ and
 $\gamma/\Omega = 0.10$, respectively.
 For $\gamma_{12}<\gamma$, both ions relax exponentially to the thermal value
 $\bar{N}$.
 When the critical condition $\gamma_{12}=\gamma$ is reached, a
 decoherence-free normal mode emerges and part of the initial excitation
 remains trapped, leading to the modified steady-state value given in
Eq.~\eqref{eq:steadyState}.}
\label{fig.expectedvalue}
\end{figure}

 Figure~\ref{fig.expectedvalue} shows the time dependence of $n_{1} (t)$ for several values of $\gamma_{12}/\gamma$, assuming the ions are initially in thermal states with mean occupation $\bar{n}_1$ and $\bar{n}_2$.  
When $\gamma_{12}<\gamma$, both ions relax exponentially to the thermal occupation $\bar{N}$, whereas if the condition $\gamma_{12}=\gamma$ is satisfied, a fraction of the initial excitations remains trapped indefinitely since
\begin{equation}
n_{j}(\infty)
= \frac{1}{2}\left(\bar{N}+\frac{n_{1}(0)+n_{2}(0)}{2}\right).
\label{eq:steadyState}
\end{equation}
 Here $n_{j}(\infty) \equiv \lim_{t\rightarrow \infty} n_{j}(t)$, and derivatives are taken before the asymptotic limit.
A short-time expansion of Eq.~\eqref{eq:population_full},  valid for
$\gamma t, \gamma_{12} t \ll 1$, further clarifies the influence of correlated dissipation,
\begin{align}
n_{j}(t)
&\approx 
n_{j}(0)\cos^{2}(\Omega t) +
n_{l}(0)\sin^{2}(\Omega t)
+ \bar{N}\gamma t
\nonumber\\
&\quad+ \frac{1}{2}
\bigg[
\frac{\gamma_{12}^{2}}{2}(n_{1}(0)+n_{2}(0))
- \bar{N}(\gamma^{2}+\gamma_{12}^{2})
\bigg]t^{2},
\end{align}
showing that cross-damping modifies the curvature of the short time population dynamics.

These results establish the dynamical regimes relevant for parameter estimation (Sec.~V) and for the generation and preservation of entanglement (Sec.~VI).

%
%

\section{Fisher information and parameter estimation}

The dynamics derived in Sec.~IV determine how the motional populations depend on
the system parameters and on the degree of reservoir correlation. We now
quantify how precisely these parameters can be inferred from energy measurements
of a single ion by evaluating the classical Fisher information \cite{frieden2004,Casella2006} associated with the parameters $\theta$ in
the occupation-number distribution $P_1(k;\boldsymbol{\theta})\equiv P_1(k)$.

Since the system remains Gaussian under linear evolution and Gaussian noise, the
reduced state of each ion is fully characterized by the local covariance matrix \cite{PhysRevA.72.012317},
\begin{equation}
\mathbf{V}_1(t)=
\begin{pmatrix}
n_1(t) & m_1(t)\\
m_1^*(t) & n_1(t)
\end{pmatrix},
\end{equation}
where $n_1(t)=\langle a_1^\dagger(t) a_1(t)\rangle$ and
$m_1(t)=-\langle a_1^2(t)\rangle$. The probability of detecting $k$ phonons ~\cite{SaNeto2022} is
then given by
\begin{equation}
P_1(k)
= \frac{{}_2F_{1}\!\left(\frac{k+1}{2},\frac{k+2}{2},1,z\right)}
{\left(1+\frac{n_1}{n_1^2-|m_1|^2}\right)^{k+1}\sqrt{n_1^2-|m_1|^2}},
\label{eq:P1_general}
\end{equation}
where ${}_2F_{1}$ denotes the Gauss hypergeometric function and
\begin{equation}
z=\frac{|m_1|^2}{\left[n_1(n_1+1)-|m_1|^2\right]^2}.
\end{equation}

For general initially uncorrelated modes one finds
\begin{equation}
\begin{aligned}
m_1 (t)= -\frac{e^{-2i\omega_0 t}}{4} \Big[ &\big(m_1(0) + m_2(0)\big)
\big( e^{2i\Omega t} e^{-(\gamma-\gamma_{12})t}
   + e^{-2i\Omega t} e^{-(\gamma+\gamma_{12})t} \big) \\
 &+ 2e^{-\gamma t} \big( m_1(0)  - m_2(0) \big) \Big],
\end{aligned}
\end{equation}
which depends on the presence of initial fluctuations in at least one of the
ions, and vanishes at long times when $\gamma_{12}<\gamma$, while at the
decoherence-free condition $\gamma_{12}=\gamma$, it approaches the asymptotic
oscillatory form
\begin{equation}
 m_1(\infty )= -\frac{ m_1(0)  +  m_2(0) }{4}\, e^{-2i(\omega_0 +\Omega)t},
\end{equation}
with a constant modulus determined by the initial second-order moments.

When both ions start in thermal states, implying zero initial fluctuations
($m_1(0)=m_2(0)=0$), Eq.~\eqref{eq:P1_general} simplifies and
reduces to the geometric distribution
\begin{equation}
P_1(k)=\frac{n_1(t)^k}{\big(1+n_1(t)\big)^{k+1}},
\label{eq:P1_thermal}
\end{equation}
so the full statistics is encoded solely in the mean occupation $n_1(t)$.

The Fisher Information Matrix (FIM) \cite{frieden2004,Casella2006} with respect to the parameter vector
$\boldsymbol{\theta}$ is 
\begin{equation}
F_{\alpha\beta}
=\sum_{k}P_1(k;\boldsymbol{\theta})
\frac{\partial\ln P_1(k;\boldsymbol{\theta})}{\partial \theta_\alpha}
\frac{\partial\ln P_1(k;\boldsymbol{\theta})}{\partial \theta_\beta},
\end{equation}
where the diagonal entries are non-negative and measure the sensitivity of the
likelihood to individual parameters, with larger values indicating higher
precision in estimation. The off-diagonal elements quantify statistical
dependencies between parameter estimates -- a positive value indicates that
increasing parameter $\theta_{\alpha}$ enhances the information available about
$\theta_{\beta}$, whereas a negative value implies that precise estimation of
one parameter reduces the available information about the other.

For the distribution~\eqref{eq:P1_thermal}, the FIM simplifies to
\begin{equation}
F_{\alpha\beta}=
\frac{1}{n_1(t)\big(1+n_1(t)\big)}
\frac{\partial n_1(t)}{\partial\theta_\alpha}
\frac{\partial n_1(t)}{\partial\theta_\beta},
\label{eq:FIM_thermal}
\end{equation}
where the parameter vector is
\begin{equation}
\boldsymbol{\theta}
=\big(n_1(0), n_2(0), \Omega, \gamma, \gamma_{12}, \bar{N}\big)^{T}\equiv \big(\theta_1, \theta_2, \theta_3, \theta_4, \theta_5, \theta_6\big)^{T},
\end{equation}

For $0\le \gamma_{12}\le \gamma$ we treat $\gamma$ and $\gamma_{12}$ as
independent parameters of the dynamical model. Consequently, the Fisher
information matrix is computed from Eq.~(44) by taking the partial derivatives
$\partial_{\theta_\alpha} n_1(t)$ with respect to each component of
$\boldsymbol{\theta}$ and only afterwards evaluating them at the chosen
numerical values of $(\gamma,\gamma_{12})$. In particular, the decoherence-free
condition is incorporated by evaluating the resulting expressions at the point
$\gamma_{12}=\gamma$ (rather than imposing $\gamma_{12}=\gamma$ prior to
differentiation). This procedure yields a well-defined Fisher matrix and allows
a consistent comparison between the fully correlated case and the partially
correlated regimes. If one instead assumes an exact constraint $\gamma_{12}=\gamma$ as part
of the model definition, then $\gamma_{12}$ is no longer an independent
coordinate and the parameter space must be reduced accordingly. In that
constrained description, it is natural to reparametrize the dissipation sector
in terms of the collective rates $\Gamma_\pm=\gamma\pm\gamma_{12}$, and in the
DFS limit $\Gamma_-=0$ is fixed while only the remaining independent
combination(s) can be estimated from data. In the remainder of this section,
however, we keep $(\gamma,\gamma_{12})$ independent and use $\gamma_{12}=\gamma$
only as an evaluation point.

\subsection{Inference of $n_1(0)$, $n_2(0)$, and $\bar{N}$}

\begin{figure}[t]
\centering
\includegraphics[width=0.5\textwidth]{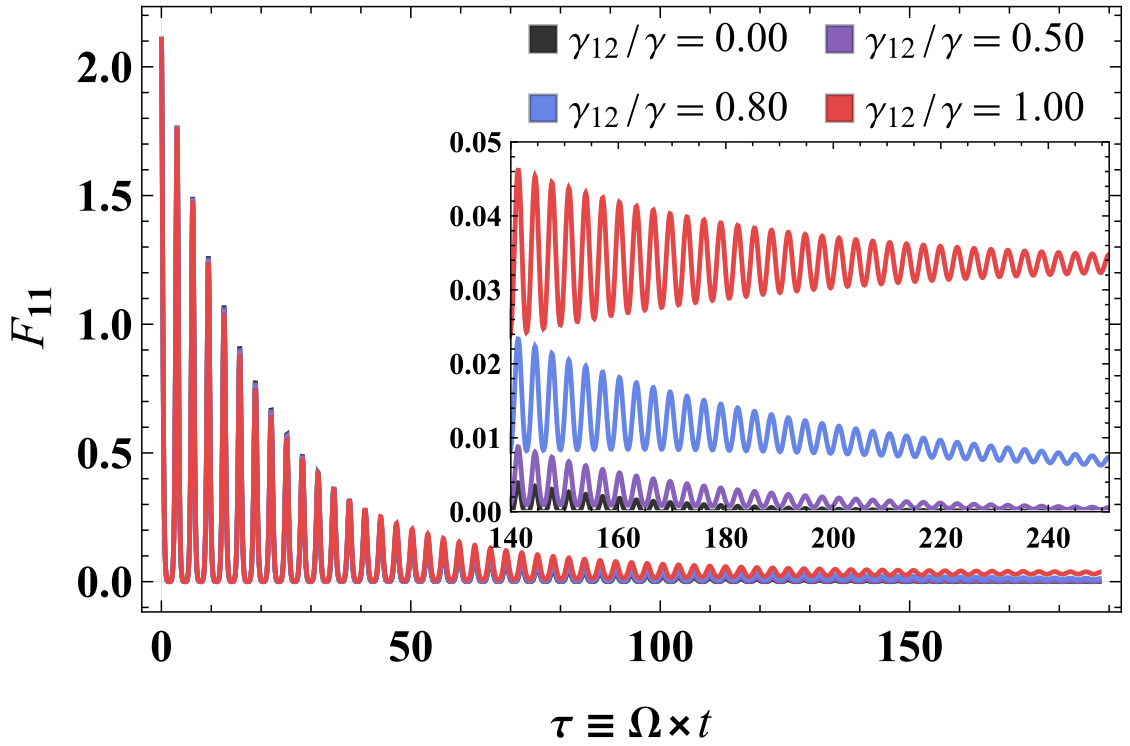}
\caption{\justifying
(Color online) Fisher information element $F_{11}$, quantifying the
sensitivity of the phonon-number distribution to the initial occupation
$n_1(0)$, for several values of the normalized cross-damping rate
$\gamma_{12}/\gamma$ with the initial conditions $ \bar{n}_{1}=0.35,\  \bar{n}_{2} = 2.3,\  \hbar\omega_{0}/k_{B}T = 1.0,\ \gamma/\Omega = 0.02$ and $ \Omega = 3.1\pi \si{\kilo \hertz}$.  The curves, from bottom to top, correspond to $\gamma_{12}/\gamma = 0.00, 0.50, 0.80, 1.00$.
For weakly correlated dissipation, $F_{11}$ displays damped Rabi oscillations
whose envelope decays at rate~$\gamma$.  
As $\gamma_{12}$ increases, the decay slows down and the short-time peak broadens.
When the fully correlated condition $\gamma_{12}=\gamma$ is reached, a
decoherence-free subspace forms and $F_{11}$ saturates to a nonzero long-time
plateau, indicating persistent information about the initial state.  The inset shows the long-time behavior toward the respective steady-state values}
\label{fig:FI11}
\end{figure}

\begin{figure}[t]
\centering
\includegraphics[width=0.5\textwidth]{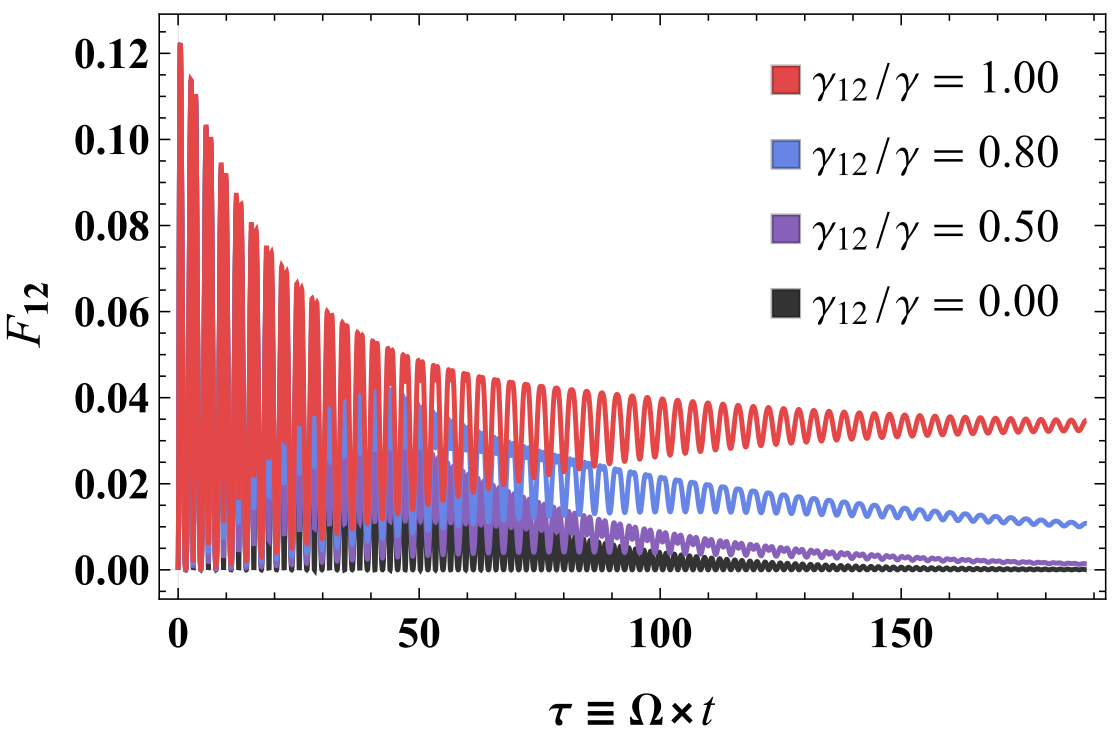}
\caption{\justifying
(Color online) Fisher information element $F_{12}$, describing correlations
between the estimators of the initial occupations $n_1(0)$ and $n_2(0)$   for several values of the normalized cross-damping rate
$\gamma_{12}/\gamma$ with the initial conditions $ \bar{n}_{1}=0.35,\  \bar{n}_{2} = 2.3,\  \hbar\omega_{0}/k_{B}T = 1.0,\ \gamma/\Omega = 0.02$ and $ \Omega = 3.1\pi \si{\kilo \hertz}$. The curves, from bottom to top, correspond to $\gamma_{12}/\gamma = 0.00, 0.50, 0.80, 1.00$.
For $\gamma_{12}=0$, $F_{12}$ oscillates around zero as excitations are
periodically exchanged between ions through the coherent coupling~$\Omega$.  
Increasing $\gamma_{12}$ leads to an asymmetric oscillation pattern and an
overall enhancement of short-time sensitivity.
In the fully correlated case $\gamma_{12}=\gamma$, a nonzero long-time plateau
emerges due to the decoherence-free subspace, indicating that correlations
between the initial occupations remain encoded in the asymptotic state.}
\label{fig:FI12}
\end{figure}

\begin{figure}[t]
\centering
\includegraphics[width=0.5\textwidth]{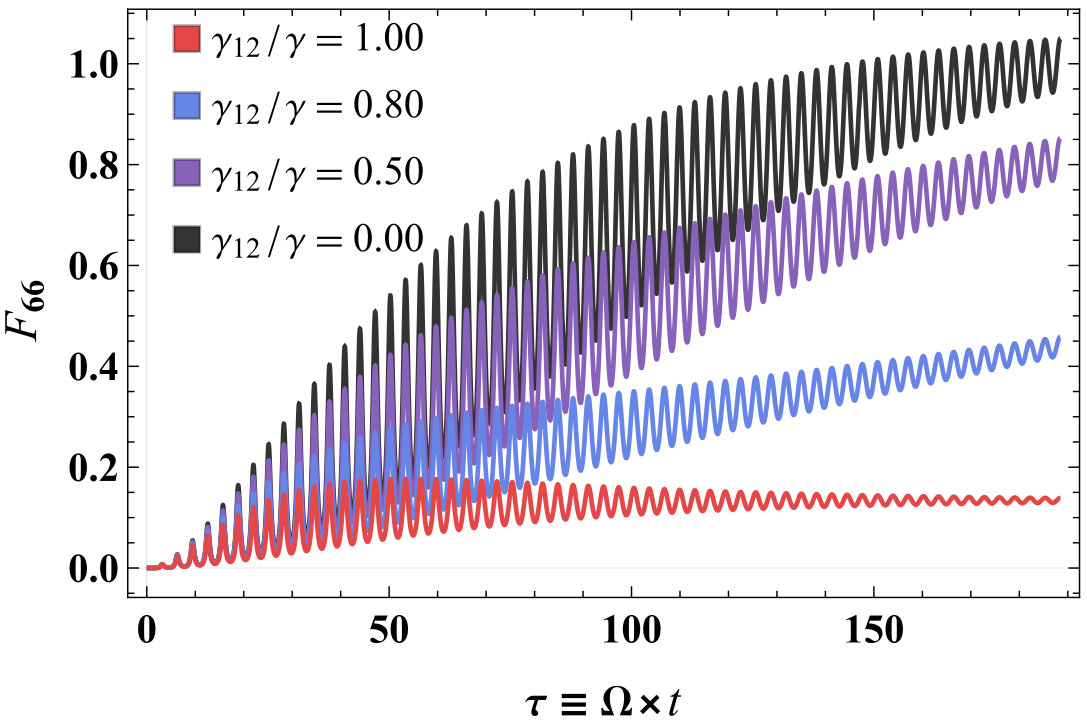}
\caption{\justifying
(Color online) Fisher information element $F_{66}$ associated with the
estimability of the reservoir thermal occupation $\bar{N}$   for several values of the normalized cross-damping rate
$\gamma_{12}/\gamma$ with the initial conditions $ \bar{n}_{1}=0.35,\  \bar{n}_{2} = 2.3,\  \hbar\omega_{0}/k_{B}T = 1.0,\ \gamma/\Omega = 0.02$ and $ \Omega = 3.1\pi \si{\kilo \hertz}$. The curves, from  top to  bottom, correspond to $\gamma_{12}/\gamma = 0.00, 0.50, 0.80, 1.00$.  
For all values of the correlation parameter $\gamma_{12}/\gamma$, $F_{66}$
monotonically increases at short times and saturates to a finite constant.
This reflects that the steady-state phonon number always carries information
about the reservoir temperature, whereas the influence of correlated
dissipation affects only the transient dynamics.  
Unlike the FI elements associated with initial occupations, $F_{66}$ does not
vanish at long times even when the bath is uncorrelated.}
\label{fig:FI66}
\end{figure}

Figures~\ref{fig:FI11}--\ref{fig:FI66} illustrate three representative elements
of the FIM: the diagonal terms $F_{11}$ and $F_{66}$ and the off-diagonal
correlation $F_{12}$. All of them display a pronounced short-time enhancement.

 Initially, when coherent evolution dominates, 
\begin{equation}
n_1(t)\simeq 
n_1(0)\cos^{2}\Omega t
+ n_2(0)\sin^{2}\Omega t
+\mathcal{O}(t),
\end{equation}
so any parameter that modifies the coherent exchange frequency or the initial
energy distribution produces a large curvature in $n_1(t)$ and hence a large FI.
This explains the rapid growth of $F_{11}$ and $F_{12}$ during the first few
Rabi periods \footnote{Similar behavior is observed by $F_{22}$ related to inference of $n_2(0)$, not shown due to symmetry with $F_{11}$.}, as seen in Figs.~\ref{fig:FI11} and~\ref{fig:FI12}.
Increasing $\gamma_{12}$ enhances this effect.
Correlated dissipation reduces the effective damping rate of the antisymmetric
collective mode $B$, which becomes subradiant and eventually decoherence free at
$\gamma_{12}=\gamma$. This slows down the overall decay of the coherent energy
exchange between the ions and increases
$\partial_{\theta_\alpha} n_1(t)$ for parameters such as $n_1(0)$, $n_2(0)$, and
$\Omega$.

The long-time limit of the FI can be understood from the stationary value of
$n_1(t)$,
\begin{equation}
n_1(\infty)=
\begin{cases}
\bar{N}, & \gamma_{12}<\gamma,\\[1mm]
\frac{1}{2}\!\left(\bar{N}+\frac{n_1(0)+n_2(0)}{2}\right), & \gamma_{12}=\gamma.
\end{cases}\label{assint}
\end{equation}
In the case $\gamma_{12}<\gamma$, the steady state is thermal, so
\begin{equation}
\frac{\partial n_1(\infty)}{\partial \theta_\alpha}=0
\qquad
\text{for all }\theta_\alpha\neq\bar{N}.
\end{equation}
Hence, by Eq.~\eqref{eq:FIM_thermal},
\begin{equation}
F_{\alpha\beta}(t\!\to\!\infty)=0,
\qquad
(\alpha,\beta)\neq(\bar{N},\bar{N}).\label{assint2}
\end{equation}
This behavior appears clearly in Fig.~\ref{fig:FI66} the FI associated with
the reservoir occupation $\bar{N}$ saturates to a finite asymptotic value,
whereas $F_{11}$ and $F_{12}$ decay to zero when $\gamma_{12}<\gamma$.

In the fully correlated case $\gamma_{12}=\gamma$, a decoherence-free subspace
(DFS) is formed, and
  \begin{equation} \frac{\partial n_1(\infty)}{\partial \theta_{\alpha}}\neq 0 \qquad
\text{for all }\theta_\alpha\neq \Omega,  \end{equation}
so part of the initial population remains indefinitely encoded in $n_1(t)$.
Thus, $F_{11}$ and $F_{12}$ acquire a nonzero long-time plateau
[see Figs.~\ref{fig:FI11} and~\ref{fig:FI12}].
In contrast, $F_{66}$ retains its usual thermal asymptote, since the steady-state
temperature is unaffected by the DFS.

The specific features relevant for inference can now be summarized.
The element $F_{11}$, shown in Fig.~\ref{fig:FI11}, quantifies the sensitivity
to the initial occupation $n_1(0)$.
For small $\gamma_{12}$ it displays damped oscillations with an envelope
decaying at the single-ion rate~$\gamma$.  
As $\gamma_{12}$ increases, the decay of the envelope slows down, and when
$\gamma_{12}=\gamma$ a clear nonzero plateau emerges at long times.
This persistent FI is a direct signature of the DFS and indicates that
measurements performed at arbitrarily late times retain information about the
initial state. By symmetry, an analogous behavior would be obtained for $F_{22}$,
associated with $n_2(0)$.

The element $F_{66}$, displayed in Fig.~\ref{fig:FI66}, describes the
sensitivity to the reservoir occupation $\bar{N}$. Unlike the FI for
initial-state parameters, $F_{66}$ envelope grows monotonically at short times and
saturates to a finite constant for all values of $\gamma_{12}$. Correlated
dissipation modifies the transient dynamics, but the asymptotic FI is almost
independent of $\gamma_{12}$, reflecting the fact that the steady state is
always thermal and that the temperature parameter remains fully estimable.

While the diagonal FIM elements are relevant for single-parameter estimation,
the off-diagonal ones are crucial for joint inference. This is exemplified by
$F_{12}$, which represents the correlation between the estimators of $n_1(0)$
and $n_2(0)$, as shown in Fig.~\ref{fig:FI12}. For $\gamma_{12}=0$, $F_{12}$
oscillates around zero because excitations periodically exchange between the two
ions. As $\gamma_{12}$ increases, the oscillations become less symmetric, and a
nonzero asymptotic value appears when $\gamma_{12}=\gamma$. This plateau
indicates that the DFS not only preserves information about each initial
occupation but also about their correlation, viewed as parameters of the
estimation problem.

\subsection{Estimation of trap and dissipation parameters: $\Omega$, $\gamma$, and $\gamma_{12}$  }

In addition to initial-state parameters and the reservoir occupation, the
Fisher information also quantifies the estimability of the intrinsic trap and
dissipation parameters -- the coherent coupling $\Omega$ (controlled by the
inter-ion Coulomb repulsion), the local damping rate $\gamma$, and the
cross-damping rate $\gamma_{12}$. These appear respectively in the FIM elements
$F_{33}$, $F_{44}$, and $F_{55}$.

\begin{figure}[t]
\centering
\includegraphics[width=0.48\textwidth]{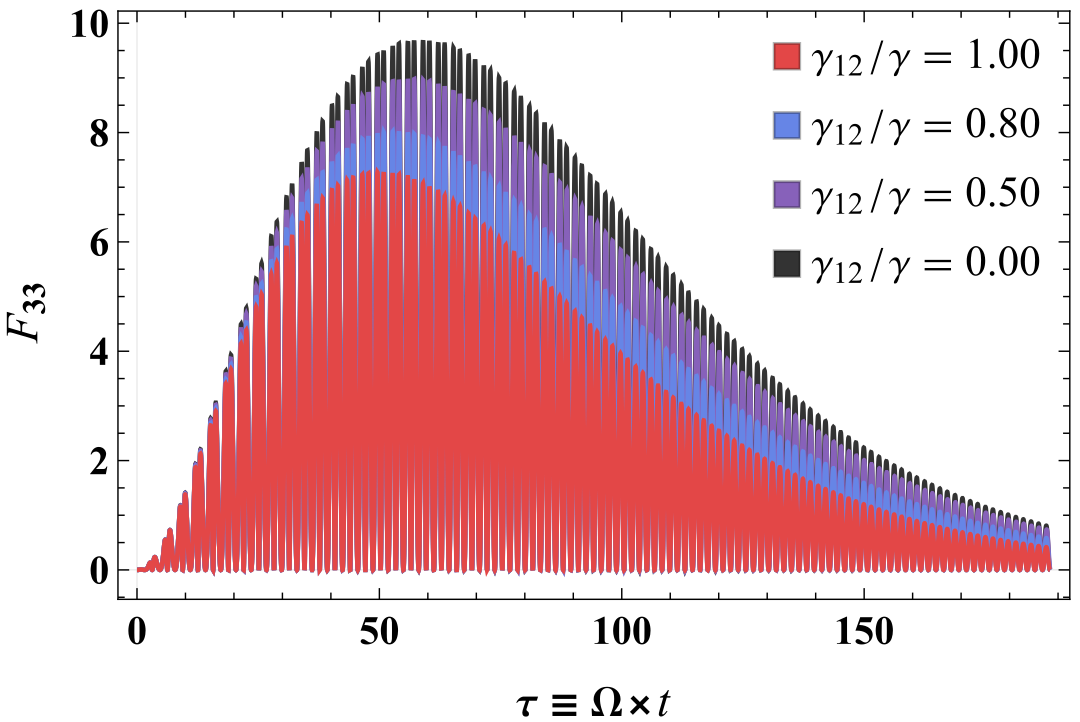}
\caption{\justifying
(Color online) Fisher information element $F_{33}$ associated with estimating
the coherent coupling $\Omega$  for several values of the normalized cross-damping rate
$\gamma_{12}/\gamma$ with the initial conditions $ \bar{n}_{1}=0.35,\  \bar{n}_{2} = 2.3,\  \hbar\omega_{0}/k_{B}T = 1.0,\ \gamma/\Omega = 0.02$ and $ \Omega = 3.1\pi \si{\kilo \hertz}$. The curves, from top to bottom, correspond to $\gamma_{12}/\gamma = 0.00, 0.50, 0.80, 1.00$.
The FI displays oscillations at short and intermediate times, reflecting the
sensitivity of $n_{1}(t)$ to the coherent exchange of excitations. Increasing
the cross-damping $\gamma_{12}$ slows the decay of the envelope and enhances
the time window over which $\Omega$ can be accurately inferred. For
$\gamma_{12}<\gamma$, $F_{33}$ vanishes at long times since the steady state
does not retain information about the coherent coupling.}
\label{fig:FI33}
\end{figure}

\begin{figure}[t]
\centering
\includegraphics[width=0.48\textwidth]{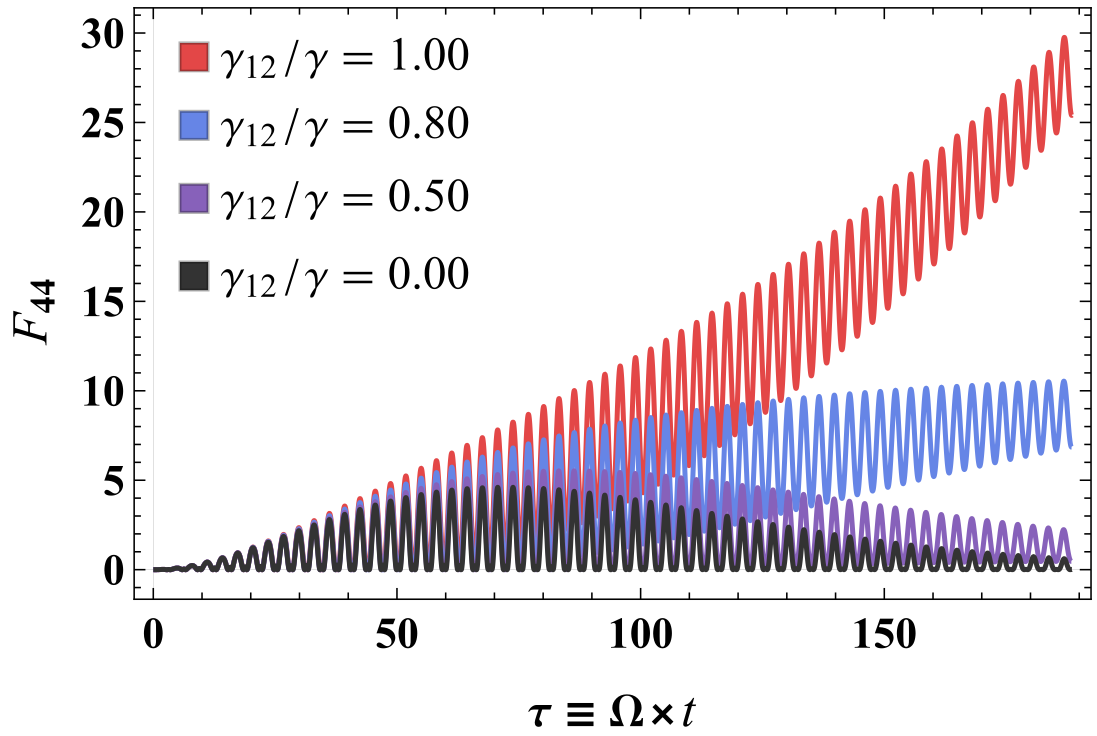}
\caption{\justifying
(Color online) Fisher information element $F_{44}$ related to the estimation of
the local damping rate $\gamma$  for several values of the normalized cross-damping rate
$\gamma_{12}/\gamma$ with the initial conditions $ \bar{n}_{1}=0.35,\  \bar{n}_{2} = 2.3,\  \hbar\omega_{0}/k_{B}T = 1.0,\ \gamma/\Omega = 0.02$ and $ \Omega = 3.1\pi \si{\kilo \hertz}$. The curves, from bottom to top  , correspond to $\gamma_{12}/\gamma = 0.00, 0.50, 0.80, 1.00$.
The FI peaks around times of order $\gamma^{-1}$, where the dissipative
contribution to the dynamics dominates over coherent exchange.  
Correlated dissipation broadens this peak and enhances short-time sensitivity.
In all cases with $\gamma_{12}<\gamma$, the FI vanishes at long times because
the steady-state population carries no information about $\gamma$.}
\label{fig:FI44}
\end{figure}

\begin{figure}[t]
\centering
\includegraphics[width=0.48\textwidth]{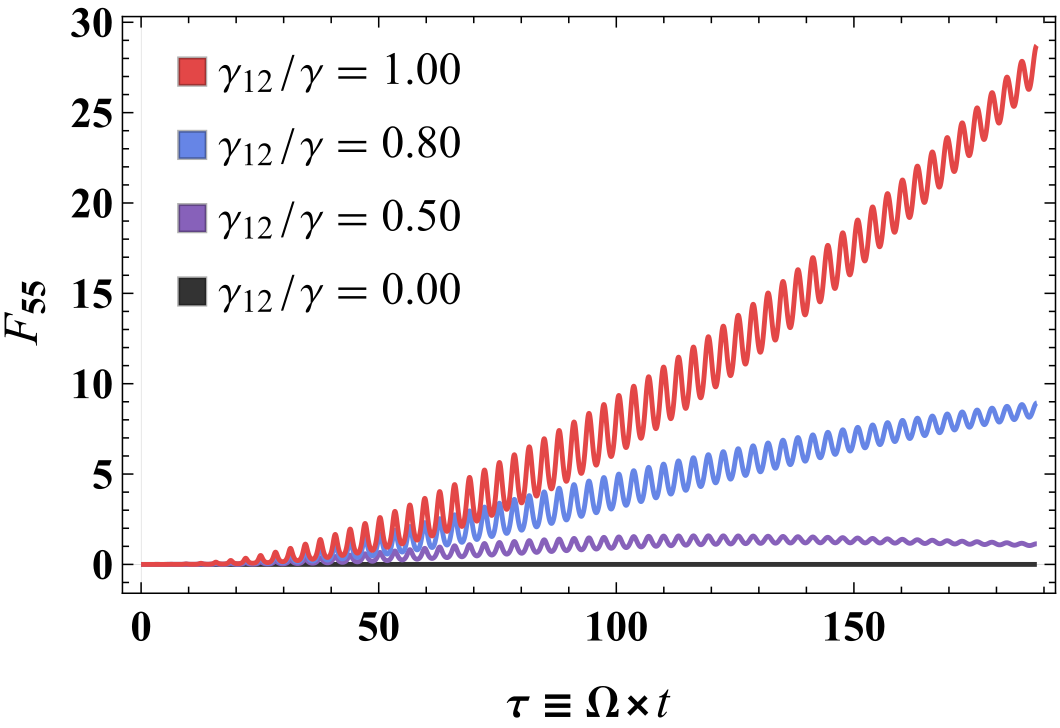}
\caption{\justifying
(Color online) Fisher information element $F_{55}$ corresponding to the estimation
of the cross-damping rate $\gamma_{12}$  for several values of the normalized cross-damping rate
$\gamma_{12}/\gamma$ with the initial conditions $ \bar{n}_{1}=0.35,\  \bar{n}_{2} = 2.3,\  \hbar\omega_{0}/k_{B}T = 1.0,\ \gamma/\Omega = 0.02$ and $ \Omega = 3.1\pi \si{\kilo \hertz}$. The curves, from bottom to top, correspond to $\gamma_{12}/\gamma = 0.00, 0.50, 0.80, 1.00$.  
The FI exhibits a strong dependence on $\gamma_{12}$: for weak cross-damping
the sensitivity is low, but it increases rapidly as $\gamma_{12}$ approaches
$\gamma$. Near the fully correlated regime, the FI develops a broad and
pronounced peak, reflecting the strong influence of $\gamma_{12}$ on the
collective decay rates. As with other transient parameters, $F_{55}$ decays to
zero at long times unless the decoherence-free condition is satisfied.}
\label{fig:FI55}
\end{figure}

Figures~\ref{fig:FI33}--\ref{fig:FI55} show the behavior of these FIM elements
for representative values of $\gamma_{12}/\gamma$. Their structure reflects the
fact that $\Omega$, $\gamma$, and $\gamma_{12}$ influence the dynamics only
through transient behavior, since none of these parameters appear explicitly in
the long-time steady state (except indirectly when the DFS is formed). We now discuss the inference of the three separate parameters.


Figure~\ref{fig:FI33} shows that the Fisher information $F_{33}$ associated with the coherent coupling $\Omega$ shows the transient coherent oscillations in the population dynamics. Increasing the cross-damping $\gamma_{12}$ does not show a significant change in the profile of $F_{33}$,  keeping the finite-time sensitivity to $\Omega$ qualitatively the same in all instances of the rate $\gamma_{12}/\gamma$. Since the steady-state population is independent of $\Omega$, $F_{33}$ vanishes asymptotically for all values of $\gamma_{12}$, including at the decoherence-free point $\gamma_{12}=\gamma$, making the estimation of $\Omega$ intrinsically a finite-time task.

The FIM element $F_{44}$, shown in Fig.~\ref{fig:FI44}, characterizes the ability
to infer the local damping rate of each ion. Since $\gamma$ governs the rate
at which excitations leak into the reservoir, $F_{44}$ peaks when the
dissipative contribution dominates over coherent exchange, typically at times
comparable to $\gamma^{-1}$.  As $\gamma_{12}$ increases, the interplay between
$\Gamma_{+}$ and $\Gamma_{-}$ modifies the decay envelope, broadening the region
of high sensitivity. Nevertheless, because $\gamma$ affects only the transient
dynamics, $F_{44}(t\to\infty)=0$ for all $\gamma_{12}<\gamma$.   
We shall talk about the  behavior of $F_{44}$ in the regime $\gamma_{12}=\gamma$ latter on.

In 
Figure~\ref{fig:FI55} $F_{55}$  displays the FIM element associated with the cross-damping.
Compared with $F_{33}$ and $F_{44}$, the FI for $\gamma_{12}$ exhibits the
strongest dependence on reservoir correlations. When $\gamma_{12}$ is small,
the effect of cross-damping is weak and the corresponding FI remains small. As
$\gamma_{12}$ increases, the FI grows rapidly -- the curvature of $n_{1}(t)$ with
respect to $\gamma_{12}$ becomes larger because this parameter directly controls
the splitting of the collective decay rates $\Gamma_{\pm}$. Near the boundary
$\gamma_{12}\to\gamma$, similar to  $F_{44}$, the FI develops a tall and broad peak, indicating that
cross-damped dynamics can be extremely sensitive to small changes in
$\gamma_{12}$. As for $F_{44}$, at $\gamma_{12}=\gamma$  the curve shows an envelope  increasing with $t^2$. 

\begin{figure}[t]
\centering
\includegraphics[width=0.48\textwidth]{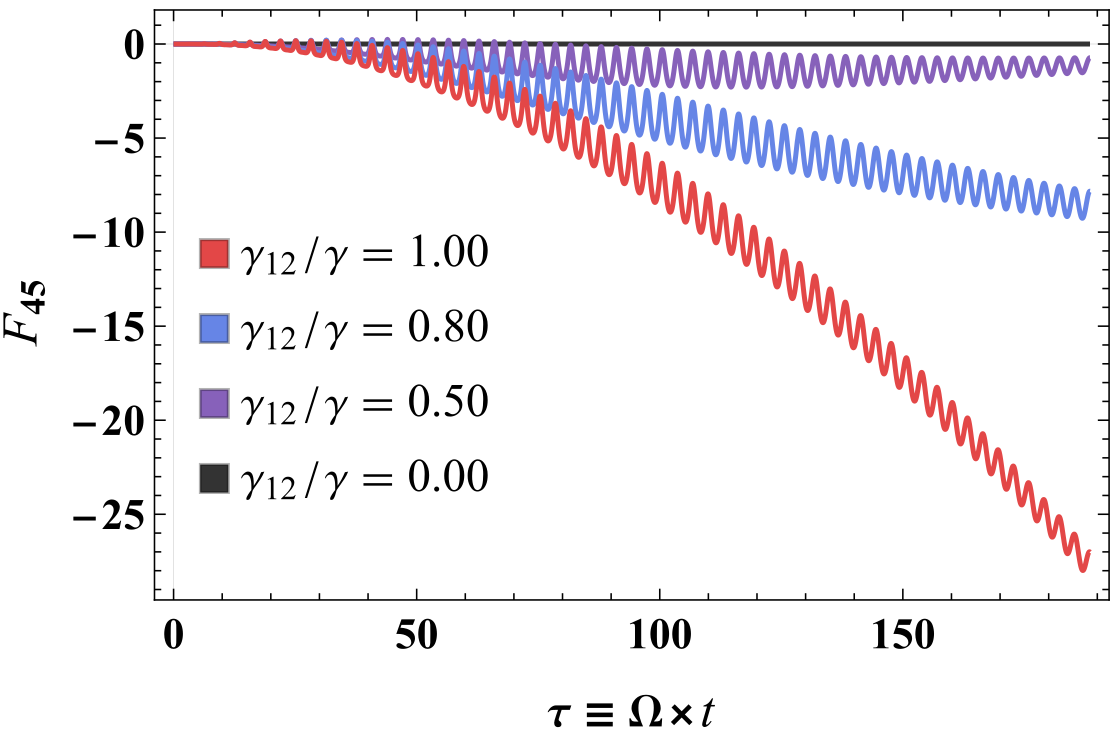}
\caption{\justifying
(Color online) Fisher information element $F_{45}$ corresponding to the simultaneous estimation
of the local, $\gamma$, and cross-damping  for several values of the normalized cross-damping rate
$\gamma_{12}/\gamma$ with the initial conditions $ \bar{n}_{1}=0.35,\  \bar{n}_{2} = 2.3,\  \hbar\omega_{0}/k_{B}T = 1.0,\ \gamma/\Omega = 0.02$ and $ \Omega = 3.1\pi \si{\kilo \hertz}$. The curves, from top to bottom , correspond to $\gamma_{12}/\gamma = 0.00, 0.50, 0.80, 1.00$ rates.}
\label{fig:FI45}
\end{figure}

The off-diagonal Fisher information element $F_{45}$ (shown in Fig. (\ref{fig:FI45})) quantifies the statistical correlation between the estimators of the local damping rate $\gamma$ and the cross-damping rate $\gamma_{12}$. Since both parameters enter the dynamics through the collective decay rates $\Gamma_{\pm}=\gamma\pm\gamma_{12}$, $F_{45}$ is generally negative, indicating an anticorrelation between their estimates. As $\gamma_{12}$ increases, this anticorrelation becomes stronger, reflecting the growing sensitivity of the dynamics to the balance between local and collective dissipation. Near the fully correlated regime $\gamma_{12}\to\gamma$, $F_{45}$ acquires a large magnitude and a broad temporal profile, signaling that the dynamics constrain primarily the difference $\gamma-\gamma_{12}$ rather than the individual rates.

This  is essential for understanding the behavior of $F_{44}$, and $F_{55}$, and, at the fully correlated limit $\gamma_{12} = \gamma$. In this regime  $\gamma_{12} =\gamma$  the curve envelope of the curve for $F_{44}$, $F_{55}$ monotonically increases with  $t^2$, since
 by taking the derivatives of (\ref{eq:population_full}) in respect to $\gamma_{12}$, or $\gamma$ and taking the limit of $t\rightarrow \infty$, leads to 
\begin{equation}
\frac{\partial n_1(\infty)}{\partial \gamma }  \sim -  \frac{\partial n_1(\infty)}{\partial \gamma_{12} } \sim   \frac{t}{2} \left[ \bar{N}- \frac{(n_{1}(0)+n_{2}(0))}{2}\right].   
\end{equation} 

Although the steady-state population in Eq.~(47) does not depend on $\gamma$ or $\gamma_{12}$ individually, the observed scaling $F_{44} = F_{55} = -F_{45} \propto t^2$ at long times indicates that the system nevertheless develops enhanced sensitivity to the equality $\gamma_{12} = \gamma$.

This apparent paradox is resolved by recognizing that while individual estimates of $\gamma$ and $\gamma_{12}$ would indeed have zero Fisher Information at infinite time, their difference $\Gamma_- = \gamma - \gamma_{12}$ becomes increasingly estimable. The quadratic growth in $F_{44}$ and $F_{55}$ reflects the system's growing ability to detect deviations from the DFS condition $\Gamma_- = 0$, even though the absolute values of $\gamma$ and $\gamma_{12}$ remain inaccessible.
Physically, this means that at long times, the system acts as an increasingly precise null detector that cannot determine what the damping rates are, but it can determine with improving accuracy whether they are equal. The equality $F_{44} = F_{55}$ and perfect anti-correlation in $F_{45}$, for large times, indicate that any change in $\gamma$ is indistinguishable from an opposite change in $\gamma_{12}$, only their difference matters for the dynamics.

This behavior illustrates a fundamental feature of estimation at parameter boundaries. FI can grow indefinitely for certain parameter combinations while decaying to zero for others. In our case, the combination $\Gamma_- = \gamma - \gamma_{12}$ acquires a Fisher Information that scales as $t^2$ at long times, allowing the equality condition $\gamma_{12} = \gamma$ to be verified with variance decreasing as $1/t^2$. Meanwhile, information about the sum $\Gamma_+ = \gamma + \gamma_{12}$ is lost in the long-time limit, reflecting that only the difference $\Gamma_-$ determines whether a DFS exists, while the absolute damping rates become irrelevant for the asymptotic state.

\subsection{Cram\'er--Rao bound and attainable precisions}

The Fisher information matrix $F_{\alpha\beta}(t)$ sets a fundamental lower
bound on the covariance matrix of any unbiased estimators
$\hat{\boldsymbol{\theta}}$ constructed from the phonon-number statistics
$P_1(k;\boldsymbol{\theta})$.
For a sufficiently large number $M$ of independent repetitions of the energy
measurement, the (classical) Cram\'er--Rao bound reads
\begin{equation}
\mathrm{Cov}\big[\hat{\boldsymbol{\theta}}\big]
\;\ge\; \frac{1}{M}\, \mathbf{F}^{-1}(t),
\label{eq:CRB_matrix}
\end{equation}
where the inequality is understood in the sense of positive semidefinite
matrices. Throughout our analysis, the parameters $\gamma$ and $\gamma_{12}$ are treated
as independent quantities, so that the Fisher information matrix and  the
associated Cram\'er--Rao bounds  remain defined even at the
decoherence-free point $\gamma_{12}=\gamma$.

For single-parameter estimation (or when the remaining
parameters are known), the variance of an unbiased estimator
$\hat{\theta}_\alpha$ satisfies
\begin{equation}
\mathrm{Var}(\hat{\theta}_\alpha)
\;\ge\; \frac{1}{M\,F_{\alpha\alpha}(t)}.
\label{eq:CRB_single}
\end{equation}

Equations~\eqref{eq:CRB_matrix} and~\eqref{eq:CRB_single} provide a direct
metrological interpretation of the Fisher-information dynamics shown in
Figs.~\ref{fig:FI11}--\ref{fig:FI66}.
The short-time enhancement of $F_{11}$ and $F_{12}$ implies that, for fixed $M$,
the minimal achievable uncertainties in the initial occupations $n_1(0)$ and
$n_2(0)$ are reduced as the cross-damping $\gamma_{12}$ increases, with the
largest gain occurring near the fully correlated regime.
Similarly, the long-time saturation of $F_{66}$ shows that the reservoir
temperature (or thermal occupation $\bar{N}$) remains estimable even after all
other parameters have become inaccessible.
In the decoherence-free regime $\gamma_{12}=\gamma$, the nonvanishing asymptotic
values of $F_{11}$ and $F_{12}$ translate into strictly positive lower bounds
for the precision of estimating the initial occupations from arbitrarily late
measurements, a feature absent for independent baths.

The same Cram\'er--Rao analysis also clarifies the attainable precision for
estimating the intrinsic trap and dissipation parameters, namely the coherent
coupling $\Omega$, the local damping rate $\gamma$, and the cross-damping rate
$\gamma_{12}$, associated with the FI elements $F_{33}$, $F_{44}$,  $F_{55}$,
respectively (see Figs.~\ref{fig:FI33}--\ref{fig:FI55}).
In all three cases, the corresponding Fisher information exhibits pronounced
peaks at short and intermediate times, reflecting the fact that these
parameters influence the dynamics only through transient behavior.
Consequently, the Cram\'er--Rao bound indicates that $\Omega$, $\gamma$, and
$\gamma_{12}$ must be inferred from measurements performed before the system
reaches its steady state.
Correlated dissipation enhances these transient FI peaks by slowing the decay
of collective modes, thereby widening the temporal window over which accurate
estimation of these parameters is possible.
However, since none of these parameters appear explicitly in the generic
long-time steady state, the corresponding FI elements vanish asymptotically
for $\gamma_{12}<\gamma$, and their estimation cannot rely on late-time data. For $\gamma_{12}=\gamma$ the asymptotic $t^2$ scaling  of $F_{44}$, $F_{55}$, and $F_{45}$  suggests that trapped ion systems near collective dissipation regimes can serve as sensitive probes for verifying decoherence-free conditions, with precision that improves linearly with measurement time, despite their inability to determine individual damping rates.

The results above can be directly mapped onto an experimentally feasible
protocol for parameter inference in trapped-ion platforms.
A typical sequence proceeds as follows:
\textit{(i).} Prepare the ions in initial thermal states with controlled
occupations $n_1(0)$ and $n_2(0)$ (or in squeezed--thermal states for
entanglement studies, as discussed in Sec.~VI).
\textit{(ii).} Allow the ions to evolve for a chosen interrogation time $t$,
during which the combined action of coherent coupling $\Omega$, local damping
$\gamma$, and cross-damping $\gamma_{12}$ shapes the motional populations.
\textit{(iii).} Perform phonon-number--resolved fluorescence detection on ion~1,
yielding a sample $\{k_1,k_2,\dots,k_M\}$ drawn from the distribution
$P_1(k;t)$ in Eq.~\eqref{eq:P1_thermal}.
\textit{(iv).} Construct unbiased estimators $\hat{\theta}_\alpha$ for the
parameters of interest, with attainable precision bounded by the
Cram\'er--Rao inequality~\eqref{eq:CRB_single}.

The time dependence of the Fisher information determines the optimal measurement
strategy.
Early-time measurements are optimal for estimating $\Omega$, $\gamma$,
$\gamma_{12}$, $n_1(0)$, and $n_2(0)$, whereas late-time measurements provide
maximal sensitivity to the reservoir occupation $\bar{N}$.
In the decoherence-free regime $\gamma_{12}=\gamma$, long-time measurements also
retain finite information about the initial occupations, offering enhanced
robustness against decoherence.

Finally, three qualitatively distinct estimation regimes can be identified.
For independent baths ($\gamma_{12}=0$), local damping causes all FI
elements except $F_{66}$ to vanish at long times, so that only the bath
temperature remains asymptotically estimable.
For partially correlated baths ($0<\gamma_{12}<\gamma$), the FI is substantially
enhanced at short times—particularly for the initial-state and trap
parameters—but all information about these quantities is eventually lost as the
system relaxes to a thermal steady state.
In the fully correlated case ($\gamma_{12}=\gamma$), a decoherence-free subspace
is formed with the long-time FI associated with $n_1(0)$, $n_2(0)$, and their
correlations remains finite due to incomplete relaxation of the antisymmetric
mode, whereas the FI for parameters that enter only through transient dynamics,
such as $\Omega$, $\gamma$, and $\gamma_{12}$ itself, still decays to zero and
must be accessed through short-time measurements.
Correlated dissipation therefore not only enhances parameter sensitivity in the
transient regime but can also preserve information about selected parameters
indefinitely, defining clear operating conditions under which dissipative
trapped-ion systems function as robust probes of both system and reservoir
properties.

Remark that we have focused on the classical Fisher information associated with
phonon-number measurements, which constitute a natural and experimentally
accessible observable in trapped-ion systems.  
The corresponding quantum Fisher information (QFI), which quantifies the ultimate
precision attainable with optimal measurements, generally exceeds or equals the
classical FI considered here.  
A full QFI analysis would require diagonalizing the Gaussian states generated by
the cross-damped dynamics and optimizing over all physically allowed POVMs, which
is beyond the scope of the present investigation.  
Nevertheless, the classical FI already captures the key qualitative features:
the enhancement of short-time sensitivity with increasing reservoir correlation,
the privileged estimability of $\bar{N}$ at long times, and the preservation of
initial-state information in the decoherence-free regime.

\section{Entanglement generation and preservation}

Correlated dissipation also has important consequences for the development and 
robustness of nonclassical correlations between the motional modes.  
In this section we analyze how the cross-damping rate $\gamma_{12}$ affects the 
creation, protection, and lifetime of Gaussian entanglement, focusing on 
initially thermal--squeezed product states and applying the PPT criterion for 
Gaussian states~\cite{de2005characterization,simon}. The Gaussian-state framework and non-classicality criteria employed here build on previous analyses of trapped-ion motional dynamics in the absence of dissipation, where squeezing was identified as a key non-classical resource \cite{AvalosNJP2025}.

Because the global state remains Gaussian for all times, its entanglement 
properties are fully determined by the covariance matrix
\begin{equation}
\mathbf{V}(t)
=
\begin{pmatrix}
\mathbf{V}_1(t) & \mathbf{C}(t)\\[1mm]
\mathbf{C}^\dagger(t) & \mathbf{V}_2(t)
\end{pmatrix}.
\end{equation}
The Peres--Horodecki PPT criterion reduces to a positivity condition for the 
partially transposed covariance matrix.  
Following Refs.~\cite{de2005characterization,simon, PhysRevA.72.012317}, we quantify entanglement 
via the symplectic invariant
\begin{equation}
Y(t)=I_1 I_2 + \left(\tfrac{1}{4}-|I_3|\right)^2 - I_4 
-\tfrac{1}{4}\big(I_1+I_2\big),
\end{equation}
where $I_1=\det\mathbf{V}_1$, $I_2=\det\mathbf{V}_2$, 
$I_3=\det\mathbf{C}$, and $I_4$ is a fourth-order invariant.  
The condition $Y(t)<0$ is a necessary and sufficient condition to the presence of entanglement for Gaussian states.  P-representable (Classical) states on each initial vibrational mode will not get entangled due to the nature of the direct interaction or the indirect interaction via the common thermal reservoir \cite{de2005characterization,MCO1}. Therefore, we consider ion 1 initially in a thermal state with $n_1(0) = \bar{n}_1$, $m_1(0) = 0$ and ion 2 in a squeezed thermal state with squeezing parameter $r$ and phase $\theta=0$, so that $n_2(0) = \left( \bar{n}_2 + 1/2 \right)\cosh(2r) - 1/2$, and $m_2(0) = \left( \bar{n}_2 + 1/2 \right) \sinh(2r)$. This choice allows us to study how initial squeezing influences entanglement generation under correlated dissipation.  
Since $\mathbf{C}(0)=0$, any entanglement must be generated dynamically—either 
by the coherent coupling $\Omega$ or by the correlated reservoir. Now, a thermal squeezed state can be non-classical, depending on the balance between $r$ and the initial thermal population of mode 2, and the direct or indirect (via reservoir) interaction will be able to entangle the ions vibrational modes.

\begin{figure}[t]
  \centering
  \begin{subfigure}[t]{0.43\textwidth}
    \centering
    \begin{tikzpicture}
      \node[inner sep=0pt] (imgA) {\includegraphics[width=\linewidth]{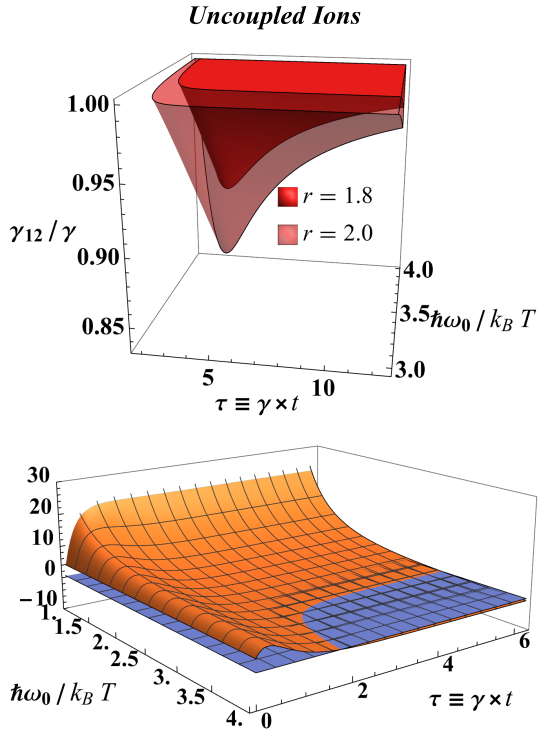}};
      \node[anchor=north west, font=\bfseries, xshift=4pt, yshift=-4pt] at (imgA.north west) {(a)};
    \end{tikzpicture}
  \end{subfigure}
  \hfill
  \begin{subfigure}[t]{0.515\textwidth}
    \centering
    \begin{tikzpicture}
      \node[inner sep=0pt] (imgB) {\includegraphics[width=\linewidth]{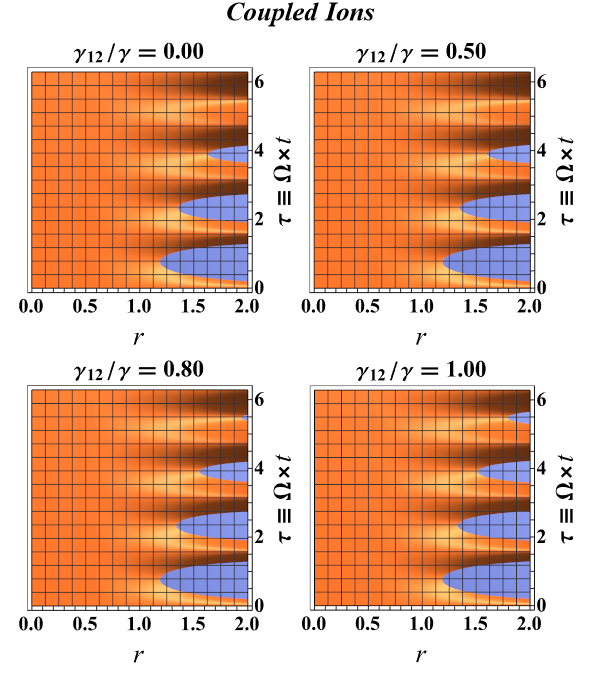}};
      \node[anchor=north west, font=\bfseries, xshift=4pt, yshift=-4pt] at (imgB.north west) {(b)};
    \end{tikzpicture}
  \end{subfigure}
  \caption{\justifying
  Separability function $Y(\tau)$ for Gaussian states of the two motional modes.
  (a) Parameter region for reservoir-induced entanglement in the absence of 
  coherent coupling ($\Omega=0$). The upper 3D surface shows the domain in the 
  space $(\tau,\gamma_{12}/\gamma,\hbar\omega_{0}/k_B T)$ where $Y(\tau)<0$, while 
  the lower panel displays the time dependence of $Y(\tau)$ for a fixed squeezing 
  parameter  $r=2.0$ and $\gamma_{12} = \gamma$, with negative values signaling entanglement. 
  (b) Entanglement dynamics for coherently coupled ions ($\Omega\neq 0$), shown 
  as density plots of $Y(\tau)$ as a function of time and squeezing $r$ for 
  several values of the normalized cross-damping $\gamma_{12}/\gamma$ (increasing 
  from the upper-left to the lower-right panel of the composite image). 
  Blue (negative) regions correspond to entangled states, illustrating how 
  increasing reservoir correlations enlarge and prolong the time windows where 
  $Y(\tau)<0$, with the longest-lived entanglement obtained near the fully 
  correlated regime $\gamma_{12}=\gamma$.}
  \label{fig.separability}
\end{figure}

Figure~\ref{fig.separability}(a) summarizes the case of uncoupled ions 
($\Omega=0$), where the common reservoir is the only mechanism capable of creating 
correlations.  
The upper 3D surface in Fig.~\ref{fig.separability}(a) shows the region in the 
parameter space $(\tau,\gamma_{12}/\gamma,\hbar\omega_{0}/k_B T)$ for which 
$Y(\tau)<0$.  
Entanglement appears only when the cross-damping is sufficiently  strong. In particular, as
$\gamma_{12}/\gamma$ approaches unity, a transient domain with $Y<0$ opens up and 
widens, revealing genuinely reservoir-induced entanglement.  
This entanglement is very sensitive to temperature.  
Larger values of $\hbar\omega_{0}/k_B T$ (lower thermal occupation) enlarge the 
negative-$Y$ region, whereas higher temperatures quickly suppress it.  

The lower panel of Fig.~\ref{fig.separability}(a) shows the corresponding time 
dependence of $Y(\tau)$ for a fixed squeezing parameter.  
At very early times, $Y>0$ because the ions are still essentially uncorrelated.  
As the common dissipation builds up correlations, $Y(\tau)$ becomes negative over 
a finite time window.  
At longer times, local heating dominates and $Y$ returns to positive values.  
For $\gamma_{12}/\gamma\simeq 1$, this negative window is substantially extended, 
showing that strongly correlated reservoirs not only generate entanglement but also 
delay its degradation even in the absence of any coherent coupling.

When the ions are coupled through the exchange Hamiltonian ($\Omega\neq0$), the 
entanglement dynamics change qualitatively.  
Figure~\ref{fig.separability}(b) displays density plots of $Y(\tau)$ as a function 
of time and squeezing $r$ for several values of $\gamma_{12}/\gamma$, corresponding 
to the four panels shown.  
For $\gamma_{12}=0$ [upper-left panel of Fig.~\ref{fig.separability}(b)], 
entanglement is generated at early times due to the beam-splitter mixing of the 
squeezed mode into the thermal one.  
The blue regions ($Y<0$) indicate intervals during which the coherent transfer of 
squeezing leads to a bipartite nonclassical state. As $\gamma_{12}$ increases to intermediate values [upper-right and lower-left panels of Fig.~\ref{fig.separability}(b)], two trends become apparent: the first 
entanglement window becomes broader, and the subsequent revivals decay more 
slowly. This behavior reflects the fact that reservoir correlations modify the collective 
decay rates of the normal modes, partially protecting the nonclassical 
fluctuations stored in them. In the fully correlated case $\gamma_{12}=\gamma$ [lower-right panel of 
Fig.~\ref{fig.separability}(b)], the entanglement windows persist for 
significantly longer times. Here the antisymmetric normal mode $B$ becomes decoherence free and retains a 
finite fraction of the initial squeezing indefinitely.  
The correlated reservoir thus plays a protective role; it reduces the effective 
damping of the entangled mode and slows the loss of quantum correlations, even 
though the local motional states of the ions continue to thermalize.

The comparison between Fig.~\ref{fig.separability}(a) and 
Fig.~\ref{fig.separability}(b) highlights the different roles played by coherent 
coupling and correlated dissipation.  
For uncoupled ions ($\Omega=0$), entanglement is generated purely by the common 
reservoir, requires strong cross-damping, and survives only within a transient 
time window before thermal noise dominates.  
For coupled ions ($\Omega\neq0$), entanglement is created mainly by coherent 
squeezing transfer, while the correlated reservoir acts as a stabilizing 
mechanism that enhances the size and duration of the entangled regions.  
In the fully correlated regime $\gamma_{12}=\gamma$, the emergence of a 
decoherence-free subspace provides the strongest enhancement and leads to 
long-lived entanglement.  
These results confirm that cross-correlated reservoirs can be exploited as a 
resource for engineering and protecting nonclassical motional states in 
trapped-ion systems.

Remark that although our separability analysis relies on the sign of the invariant
$Y(\tau)$ to detect entanglement, its amount is also reflected in the
Minkowski structure of $Y$, as established in Ref.~\cite{PhysRevA.88.052324}.
In that formulation, the negativity of $Y$ is not only a binary indicator of
nonseparability but also quantifies, through its depth, the distance of the
state from the boundary of the separable set in the associated pseudo-Riemannian
geometry.  
Thus, the separability surfaces shown in Fig.~\ref{fig.separability} capture not
only the onset of entanglement but also, through the magnitude of $Y$, the
geometric robustness of the generated quantum correlations in the sense of the
Minkowski metric structure of Gaussian-state entanglement.

\section{Conclusions}

We have investigated the dynamics of two trapped ions interacting with a common
thermal reservoir, focusing on how cross-correlated dissipation modifies
heating, relaxation, parameter sensitivity, and entanglement generation.
Starting from a microscopic system--reservoir model, we derived the corresponding
Heisenberg--Langevin equations and identified the emergence of collective decay
channels with rates $\Gamma_{\pm}=\gamma\pm\gamma_{12}$. When the cross-damping
reaches its maximal value $\gamma_{12}=\gamma$, the antisymmetric normal mode
becomes decoherence free, allowing a fraction of the initial motional population
to persist indefinitely.
The altered decay structure has direct implications for metrology. Using the
Fisher information associated with phonon-number measurements, we showed that
reservoir correlations enhance short-time sensitivity to both system and bath
parameters by slowing the decay of coherent energy exchange. For
$\gamma_{12}<\gamma$, all information except that associated with the reservoir
temperature is lost at long times, whereas in the fully correlated regime the
decoherence-free subspace preserves a finite amount of information about the
initial motional populations. Thus, correlated dissipation can improve transient
estimation accuracy and, in special regimes, prevent the complete asymptotic
loss of selected parameters.
The entanglement analysis further illustrates the role of the common reservoir.
For uncoupled ions, cross-damping alone can generate entanglement, though only
under strong reservoir correlations and sufficiently low temperatures. When
coherent coupling is present, reservoir correlations protect part of the
nonclassical fluctuations exchanged between the ions, extending entanglement
lifetimes; at $\gamma_{12}=\gamma$, the decoherence-free mode yields particularly
long-lived entanglement windows.

The parameter regimes explored in this work are compatible with current trapped-ion experiments. Typical axial trap frequencies in the range $\omega_0/2\pi \sim 0.1$--$2\,\mathrm{MHz}$, coherent coupling strengths $\Omega/2\pi$ of a few kHz, and heating rates $\gamma/2\pi$ from a few Hz to tens of Hz are routinely reported. Correlated dissipation naturally arises from shared electrodes, common voltage sources, and finite correlation lengths of surface electric-field noise, and can be further engineered using controlled noise injection or tailored reservoir engineering schemes. In this context, the Fisher-information analysis presented here provides direct guidance for optimal interrogation times and parameter regimes: transient measurements maximize sensitivity to $\Omega$, $\gamma$, and $\gamma_{12}$, while late-time measurements allow verification of decoherence-free conditions through the enhanced sensitivity to the difference $\Gamma_-=\gamma-\gamma_{12}$. These features suggest that correlated dissipation can be exploited not only as a source of noise, but also as a diagnostic and metrological resource for characterizing dissipation mechanisms and verifying collective decoherence suppression in trapped-ion platforms. The framework developed here provides practical guidance for
exploiting common-reservoir effects in trapped-ion systems and identifies
operating regimes in which dissipation can serve as a controlled resource for
precision sensing, reservoir engineering, and quantum-state manipulation.

\section*{Acknowledgments}

CFPA acknowledge the support from the Coordenação de Aperfeiçoamento de Pessoal de Nível Superior -- Brasil (CAPES) -- Finance Code 001. 
MCO acknowledges partial financial support from the National Institute of Science and Technology for Applied Quantum Computing through CNPq process No. 408884/2024-0 and by FAPESP, through the Center for Research and Innovation on Smart and Quantum Materials (CRISQuaM) process No. 2013/07276-1.

\nocite{*}
\bibliography{bibliography}
\end{document}